\documentclass[12pt,fleqn]{article}
\usepackage{amsmath}
\usepackage{amsfonts}
\usepackage{amssymb}
\usepackage{graphics,graphicx}
\usepackage{cite}

\begin{document}

\title{Role of initial conditions in the Universe formation}
\author{S.G. Rubin \\
National Research Nuclear University "MEPhI", \\ (Moscow Engineering Physics Institute)  \\
sergeirubin@list.ru}

\maketitle

\begin{abstract}
A dependence of low energy physics on initial conditions in the framework of multidimensional gravity is discussed. It is shown that the observable symmetries could be a result of specific topologies originated from space-time foam.
\end{abstract}

\section{Introduction}

Progress made in the fundamental physics, like the Higgs boson discovery, the idea of the inflationary period shows the correctness of physics development. Nevertheless a large number of theories and approaches indicates that there is  problem in description of observational and experimental data based on a unified approach.


The aim of this paper is to analyze ways of low energy physics
formation - the origination of the observable symmetries and the
physical parameters. The latter is tightly connected to the fine tuning problem - the observational fact
of narrowness of allowed physical parameters range, see e.g. \cite{fine}. The discussion is performed on the basis of the extra space concept. Despite the absence of its direct experimental confirmations, the extra space allows one to effectively explain a considerable number of phenomena and indicates the direction of their theoretical investigation. It also permits to clarify some problems of modern cosmology and the Standard Model that are discussed in terms of extra-dimensional gravity \cite{1998PhLB..429..263A,1999NuPhB.537...47D,
1999PhRvL..83.3370R,
2002PhRvD..65b4032A,2004PhRvD..70a5012C,2005PhRvD..71c5015C,2007CQGra..24.1261B}.
Additional postulates like existence of strings, branes and
multidimensional black holes open new possibilities in
description of various effects. Nevertheless many
issues can be considered without their inclusion as primary postulates. In this paper the ways based on pure extra space concept are discussed. More definitely we accept the following:

\begin{list}{}{} 
\item - random initial conditions  - a metric and topology of a manifold appeared from the space-time foam - play the key role. They lead to rich physics whose variety is connected with a huge number of stationary metrics of an extra space. The probability of the Universe birth was calculated in different approaches with the results
differing radically from one another \cite{Vilenkin}. It is not strange due to the absence of the theory of
Quantum Gravity. In this study it will be enough
to suppose nonzero probability of any metric originated from space-time foam. As was shown in \cite{2006PhRvD..73l4019B} some of
these metrics evolve classically to stationary states.

\item - a compact extra space with sufficiently large dimensionality. The number of extra dimensions has long been a matter of dispute. For example, the Kaluza-Klein model originally contained one extra dimension. At present, infinite-dimensional spaces and even variable-dimensional spaces \cite{Castro,Bleyer} are being discussed. In this paper the dimensionality is not fixed.
\item - Lagrangian is a nonlinear function of the Ricci scalar and other invariants and contains no any fields other than metric. This assumption permits to stabilize an extra space size \cite{2007CQGra..24.1261B}.
\end{list}

These issues lead to significant consequences and some of them are discussed in this paper. The first one is the origin of the low energy symmetries. It is known that
each symmetry of extra space corresponds to specific observable symmetry. At the same time symmetric extra spaces are hardly produced from space-time foam. In the paper \cite{2008PhRvD..77k5012P}  the symmetries formation mechanism related to the entropy flow from the extra space to the main one was
elaborated. Due to the entropy decrease in the compact subspace, its
metric undergoes the process of symmetrization during some period
after its quantum nucleation. Symmetry restoration relaxation time
depends on many issues and could overcome the inflationary period. A
question that remains is why the extra space entropy   is not
decreased until a widest symmetry is restored? How could one explain
the existence of observable symmetries like $SU(2)$ and $U(1)$ in
this framework? In this paper the way to solve the problem is discussed.

Another question is the origination of the physical parameters and their specific values. Section \ref{ways} is devoted to elaboration of mechanisms responsible for various low energy physical parameters. In this connection, new class of stationary metrics discussed in subsection \ref{point} plays an important role.

Analytical calculations become much easier and one could move far ahead supposing that a curvature of extra space is much larger than a curvature of our 4-dim space
\begin{equation}\label{ll}
R_4 \ll R_n
\end{equation}
This looks natural for the size  $l_n < 10^{-18}cm$ of compact extra space as compared to the Schwarzschild radius $r_g \sim 10^6cm$ of stellar mass black hole where the largest curvature in the modern Universe can occur.

\section{Separation of an extra space from the main one}

Starting from this Section it is implied that a characteristic scale of extra space is small and its geometry has been stabilized shortly after  the Universe creation. The stabilization problem is discussed in  \cite{2007JHEP...11..096G,2012PhLB..718..237K} in the framework of multidimensional gravity with higher order derivatives.
As a common basis, consider a Riemannian manifold $M_{4+n}=M\times M'$
equipped by metric
\begin{equation}\label{metric}
ds^2 = g_{\mu\nu}(x)dx^{\mu}dx^{\nu} + G_{a\mu}(x,y)dy^adx^{\mu}+ G_{ab}(x,y)dy^a dy^b
\end{equation}
with $a, b$ running over $5,6,..,n$ and $\mu,\nu$ over $1,2,3,4$. Here $M$, $M'$ are the manifolds with metrics $g_{\mu\nu}(x)$ and $G_{ab}(x,y)$ respectively. The coordinates of the subspace $M$ are denoted as $x$; $y$ is the same for $M'$. We will refer to $4$-dim space $M$ and $n$-dim compact space $M'$ as the main space and an extra space respectively.

The off diagonal components $G_{a\mu}(x,y)$ describe 4-vector fields in the low energy limit. Their contribution is not discussed here and will be omitted in the following.
We also neglect the $y$ dependence of the metric $g_{\mu\nu}(x)$ because the warped spaces have been intensively studied during last decade.

A time behavior of the metric tensor $G_{ab}(x,y)$ is governed by the classical equations of motion and varies under a variation of initial conditions. As was shown in \cite{2012PhLB..718..237K} the energy dissipation into the main space $M$ leads to an entropy decrease of the manifold $M'$ and appearance of some friction term in the classical equations for the extra metric $G_{ab}(x,y)$. This term stabilizes the extra metric. Finally the inflationary process strongly smooths out space inhomogeneities so that
\begin{equation}\label{stabi}
G_{ab}(x,y)
\xrightarrow{t\rightarrow\infty}G_{ab} (t,y)
\end{equation}
at the modern epoch. Time dependence of the external metric was discussed in the framework of the Kaluza-Klein cosmology and Einstein's gravity \cite{Abbott}. If the gravitational Lagrangian contains nonlinear in the Ricci scalar terms the extra metric $G_{ab}$ could tend to a stationary state \cite{2006PhRvD..73l4019B,2007CQGra..24.1261B} $$G_{ab}(t,y)\rightarrow G_{ab}(y),$$
see also \cite{Carroll,Nasri} for discussion.

Let us estimate the rate of stabilization of the
extra space. Weak deviations of the geometry from an equilibrium configuration can be interpreted as excited states with the mass $m_{KK}$,(see, for example, \cite{Antoniadis}). Since this is the only scale, the decay probability is expected to satisfy the decay probability satisfies the relationship $\Gamma \sim m_{KK}\sim
1/L_d$, where $L_d$ is the characteristic size
of the extra space.  According to observations $L_d \leq 10^{-18}$cm so that the lifetime of the excited state is $t_1 \sim L_d \leq 10^{-28}$s. Therefore, the extra space reaches a stationary state long before the onset of the primordial nucleosynthesis but, possibly, after completion of the inflationary stage.


The asymptotic form of the metric
\begin{equation}\label{metric2}
ds^2 = g_{\mu\nu}(x)dx^{\mu}dx^{\nu} +  G_{ab}(y)dy^a dy^b
\end{equation}
strongly facilitates the analysis because the Ricci scalar represents the simple sum of the Ricci scalar $R_4$ of the main space and the Ricci scalar $R_n$ of the extra space
\begin{equation}\label{sum}
R=R_4 + R_n
\end{equation}
what can be easily checked analytically. The metric \eqref{metric2} is intensively studied for the case of maximally symmetric extra space \cite{Muko}

We start with the action in the form \cite{2014JCAP...01..008B,2007CQGra..24.3713S}
\begin{equation}\label{act1}
S=\frac{m_D ^{D-2}}{2}\int d^4 x d^n y \sqrt{|g(x)G(y)|}f(R);\quad f(R) = \sum\limits_k {a_k R^k }.
\end{equation}
and  arbitrary parameters $a_k$. It is supposed throughout the paper that there exist stationary solution(s) of classical equations of motion what impose some restriction on the function $f(R)$, see \eqref{eq6}.  This restriction is quite weak though linear gravity is excluded for $n\neq 2$.

Using inequality \eqref{ll} and expression \eqref{sum} the Taylor expansion of $f(R)$ in  Eq. (\eqref{act1} gives
\begin{eqnarray}\label{act2}
&&S= \frac{m_D ^{D-2}}{2}\int d^4 x d^n y \sqrt{g(x)} \sqrt{G(y)} f(R_4 + R_n ) \nonumber \\
&& \simeq \frac{m_D ^{D-2}}{2}\int d^4 x d^n y\sqrt{g(x)} \sqrt{G(y)}[ R_4(x) f' (R_n (y) ) + f(R_n (y))] \nonumber \\
&& = \int d^4x  \sqrt{g(x)}\left[\frac{M^2 _{Pl}}{2}R_4 +  \frac{m_D ^{D-2}}{2}\int d^n y \sqrt{G(y)} f(R_n)\right]
 \end{eqnarray}
where $D=n+4$. The Planck mass
\begin{equation}\label{MPl}
M^2 _{Pl}=m_{D}^{D-2}\int d^n y  \sqrt{G(y)}f' (R_n (y) )
\end{equation}
and the cosmological $\Lambda$ term
\begin{equation}\label{density}
\Lambda =-\frac{m_D ^{D-2}}{2}\int d^n y \sqrt{G(y)} f(R_n)
\end{equation}
depend on specific stationary geometry $G_{ab} (y)$ and hence on initial conditions. This point will be discussed in Section \ref{ways}.

Stationary configuration $G_{ab} (y)$ is determined by static classical equations followed from action \eqref{act2}
\begin{equation}\label{eqS1}
\frac{\delta S}{\delta G_{ab} (y)}=\frac{\delta S_1}{\delta G_{ab} (y)}=0, \quad S_1 = \frac{m_D ^{D-2}}{2}V_4 \int d^ny\sqrt{|G|  } f(R_n),
 \end{equation}
where $V_4 =\int d^4x  \sqrt{g(x)}$. Explicit form of the equations is as follows
 \begin{equation}\label{eq2}
 f'(R)R_a ^b -\frac{1}{2}f(R)\delta_a ^b -\nabla_{a}\nabla^{b}f'+\delta_a ^b\square f' =0.
 \end{equation}
If the extra space is 2-dimensional, only one equation in system \eqref{eq2} remains independent what strongly facilitates the analysis.

\section{Ways to produce different universes.}\label{ways}
It is well known that the range of admissible physical parameters must be extremely narrow (the fine tuning problem) for such a complex structure as our Universe to exist, which is difficult to explain.

This Section is devoted to description of some mechanisms that can be used to fit parameters to the observable ones. The first example is based on the well known properties of hyperbolic spaces.

\subsection{Hyperbolic space}
If a solution of the system \eqref{eq2} represents maximally symmetric metric, $R=const$, this system acquires the simple form
\begin{equation}\label{eq6}
 W\equiv \frac{n}{2}f(R) - Rf'(R)=0.
\end{equation}
where the connection between the Ricci scalar and the Ricci tensor
$R^b _{a}=\delta^b _{a}R/n$
is used.
We have postulated a complicated form of the function $f$ from the beginning so that several solutions of such a sort do exist. Each of them is realized under some set of initial metrics. Consider the set $\{G_{in}\}$ of initial metrics which leads to the solution $R=R_- =const <0$.
There is no rigid connection between the Ricci scalar and a characteristic size $L$ of a compact hyperbolic space.  This was the subject of intensive discussion in a literature, see e.g. \cite{2002PhRvD..66d5029N}

The volume of a smooth compact hyperbolic space is
\begin{equation}
  V_{2}=\int d^2 y \sqrt{G_2} \simeq \frac{2}{R_-} e^{\alpha}, \quad \alpha = (n_{eff}-1)L/r_h
\end{equation}
where $n_{eff}$ is an effective dimensionality of extra space. The Planck mass \eqref{MPl}  depends on the Ricci scalar $R_-$ and on the size $L$ of extra space,
\begin{equation}\label{MP2}
M^2 _{Pl}=m_D ^2\int d^2 y  \sqrt{G(y)}f' (R_- ) \simeq m_D ^2   \frac{f' (R_- )}{R_-} e^{ (n_{eff}-1)L/r_h}.
\end{equation}

While the curvature radius $r_h =\sqrt{2/R_-}$ of the hyperbolic compact space  depends on the parameters of the function $f$, the characteristic sizes $L$ represent countable set bounded from below at fixed $r_h$. Evidently there exists a subset $\{G_L\}\subset \{G_{in}\}$
of initial metrics which leads to the stationary metric with the same Ricci scalar $R_-$ and specific size $L$ of the extra space.
The only restriction to the extra space size is \cite{2002PhRvD..66d5029N}
\begin{equation}\label{restric}
L>r_h.
\end{equation}
One may conclude that for any value of $m_D$ there exists an initial set of metrics $\{G_L\}$ so that the Planck mass \eqref{MP2} acquires the observable value.
The set of various Planck mass provided by hyperbolic spaces is the countable one. The following discussion is devoted to a production mechanism of continuous set of universes.

\subsection{Point-like defects}\label{point}

Maximally symmetrical metrics discussed above are  particular solutions to equation \eqref{eq2} with specific boundary conditions. Other solutions are the result of  different boundary conditions. The latter in turn depend on initial conditions in analogy with the Schwarzschild metric which is the result of initial spatial distribution of energy density. The integration constant in the Schwarzschild solution is related to black hole mass gathered from a surrounding media.

Let us find a metric of compact space with point-like defect. If an extra space is 2-dimentional, only one equation in system \eqref{eq2} remains independent what strongly facilitate the analysis. Let this equation is the trace of equation \eqref{eq2}
\begin{equation}\label{eqtrace}
\square f'_R =W(R)\equiv f-f'_R R;\quad f'_R=\frac{df}{dR}, \quad n=2,
\end{equation}
where
\begin{equation}
\square =\dfrac{1}{\sqrt{G}}\partial_a \sqrt{G}G^{ab}\partial_b ; a,b= 1,2
\end{equation}

Consider the metric
\begin{equation}\label{U1metrics}
ds^2 = r(\theta)^2 (d\theta^2 + \sin^2(\theta)d\phi^2 )
\end{equation}
of a manifold with the sphere topology. In this case equation \eqref{eqtrace} acquires the form
\begin{equation}\label{eqtrace1}
\frac{1}{r(\theta)\sin(\theta)}\partial_{\theta}\left( \sin(\theta)\partial_{\theta}f'_R\right)  =W(R)
\end{equation}
which may be considered as the system of differential equations together with the expression
\begin{equation}
\label{Ricci}
R=\frac{2}{r(\theta)^4\sin(\theta)}(-r'r\cos(\theta)+r^2 \sin(\theta)+r'^2 \sin(\theta)-\sin(\theta)rr'')
\end{equation}
for  the  Ricci scalar. In expression \eqref{Ricci} prime means $d/d\theta$.

Solutions $r(\theta ;c_1,c_2,c_3,c_4)$ of the system of differential equations \eqref{eqtrace1} and \eqref{Ricci} depend  on four arbitrary integration constants $c_i$. The latter are determined by boundary conditions. As the result we have continuous set of universes characterized by the Planck mass \eqref{MPl} and the cosmological constant \eqref{density} that depend on arbitrary constants. There is maximally symmetric solution among them with the Ricci scalar being the solution of equation $W(R)=0$, see also \eqref{eq6}.

Numerical analysis of arbitrary solutions will be performed in future. An asymptotic behavior $\theta \rightarrow 0$ of inhomogeneous solutions can be obtained analytically in case of a quadratic form of the function
\begin{equation}\label{fR}
f(R)=u_1 (R-R_0 )^2 +u_2 .
\end{equation}
Let the asymptotic solution has the form
\begin{equation}\label{rtet}
r(\theta)=\frac{C}{\theta^b}, \quad b>0,\quad\theta \rightarrow 0.
\end{equation}
Direct substitution of expression \eqref{rtet} into \eqref{Ricci} gives
\begin{equation}
R(\theta)=\frac{2}{C^2}(1-b/3)\theta^{2b}\quad \theta \rightarrow 0
\end{equation}
As a result, equation \eqref{eqtrace1} aquires simple form
\begin{equation}
16u_1 C^{-4}b^2 (1-b/3)\theta^{4b-2}=u_2 +u_1 R_0 ^2
\end{equation}
so that both parameters $b$ and $C$ are known
\begin{equation}\label{parbc}
b=\frac{1}{2}, \quad C=\left( \frac{3}{10}(u_2 /u_1 +R_0 ^2)\right) ^{-1/4} .
\end{equation}
Thus solution
\begin{equation}\label{asymp}
r(\theta)=\frac{C}{\theta ^{1/2}}; \quad \theta \rightarrow 0.
\end{equation}
describes a class of metrics with different boundary conditions and the same asymptotics. This point-like defect has no singularity at $\theta =0$ due to smooth behavior of the Ricci scalar
\begin{equation}\label{Ricci3}
R(\theta )=\frac{5}{3C^2}\theta .
\end{equation}
It is important to note that metric \eqref{U1metrics}, \eqref{asymp} possesses $U(1)$ symmetry what leads to observational 4-dimensional gauge field in full compliance with the original Kaluza-Klein model. This subject is also considered in the following subsection.

\subsection{Formation of gauge symmetries}

In the framework of multidimensional approach the existence of gauge symmetries is the result of  an extra space isometries \cite{1997PhR...283..303O,Blagojevic,2008IJMPD..17..785C}.  The remaining problem sounds as "why an extra space possesses symmetries"?  There is no reason to assume that the geometry or/and topology of extra space is simple if one takes into account the quantum origin of the space itself due to metric fluctuations in the space-time foam. Moreover it seems obvious that a measure $\mathfrak{M}$ of all symmetrical spaces equals zero so that the probability of their origination $\mathcal P=0$. Hence some period of extra space symmetrization had to exist \cite{2001PhRvD..63j3511S,2009JETP..109..961R,2002PhRvD..66b4036C,2003Ap&SS.283..679G,2006PhRvD..73l4019B,2012PhLB..718..237K}.
In the paper \cite{2012PhLB..718..237K} the entropic mechanism of space symmetrization after its formation is developed. It was shown that the stabilization of extra space and its symmetrization occur simultaneously. This process is accompanied by a decrease in the entropy of an extra space and an increase in the entropy of the main one.

Nevertheless one puzzle remains unsolved. Indeed, if the entropy reasons are responsible for an extra space symmetrization then all observed symmetries must relate to maximally symmetrical extra spaces  \cite{2012PhLB..718..237K,2013arXiv1311.0220C}. This contradicts the observations. How could e.g. $U(1)$ symmetry, necessary ingredient of the Standard Model, be formed in this approach? Its appearance would be natural for one-dimensional extra space, but the latter can not be stabilized in a natural way.

The problem becomes more clear if one notices that maximally symmetrical metrics are solutions to equation \eqref{eq6} which is a particular case of equation \eqref{eqtrace}. Previous subsection was devoted to the description of stationary but nonuniform in the Ricci scalar solution to equation \eqref{U1metrics}, \eqref{asymp}, \eqref{Ricci3} which possess $U(1)$ symmetry only. As a result, a final symmetry depends on initial conditions and hence a manifold is not necessarily maximally symmetric.

Up to now we have not involved topological arguments that are crucial for the following consideration. A geometry describing a stationary metric depends on both boundary conditions and a space topology. The same is true for isometries of manifolds (the number of the Killing vectors equals one and three for a torus and a sphere correspondingly).

A stationary extra space of genius 1 (e.g. torus), studied below as an example,  contains areas both with positive and negative curvature. This imposes certain conditions on initial Lagrangian. Let the function $f$ in \eqref{act1} be a fourth order polynomial in the Ricci scalar $R$ written in the form
\begin{equation}\label{f}
f(R)=u_1 + \frac{u_2}{4}[(R-R_0)^2 -R_1 ^2]^2+u_3 R.
 \end{equation}
Function $f$ has two minima provided that parameter $u_3$ is not very big.


Approximate solution to \eqref{eq2} can be found by direct minimization of action \eqref{eqS1} on a limited class of metrics. More definitely, consider the torus metric with the interval

\begin{equation}\label{torus1}
ds^2 = g_{\mu\nu}(x)dx^{\mu}dx^{\nu} -  b^2 d\theta ^2 - (a+bcos\theta)^2 d\phi ^2
\end{equation}
Then the action $S_1$ in \eqref{eqS1} is some function of two parameters $a,b$ and has the form
\begin{equation}\label{act3}
S_1(a,b)=\frac{V_4}{2}\int d\phi d\theta \cdot b\,(a+b\, cos\theta )f(R_{torus} )
\end{equation}
in $m_D$ units.
A function to be minimized is obtained by substitution the Ricci scalar
\begin{equation}\label{Rtor}
R_{torus}=\frac{2\cos(\theta)}{b(a+b\cos(\theta))}
\end{equation}
and function \eqref{f} into action \eqref{act3}.

Numerical minimization indicates that a solution does exist in wide range of initial parameters. For example the solution
\begin{equation}\label{par}
a=10.23,\, b= 4.17
\end{equation}
takes place for the parameter values
\begin{equation}\label{param}
u_1 = -0.5\cdot 10^{-5}; u_2 = 100; u_3= -0.5\cdot 10^{-4}; R0 = -0.02; u = 0.05.
\end{equation}
The Planck mass \eqref{MPl} equals $M_{P} =0.9 m_D ^2$  in this case.

Thus a surface of genius 1 is able to evolve to the usual stationary torus with metrics \eqref{torus1} provided that function $f(R)$ is quite complicated. The presence of the Killing vector produces $U(1)$ gauge symmetry of low-energy theory in our 4-dim space-time. As the result, there are no other Killing vectors - topology terminates the entropy flow.

As the preliminary conclusion, it is shown that the observable symmetries arise at early stage if necessary conditions take place. Firstly, Lagrangian \eqref{act1} should have several minima, secondly, initial topology of an extra space originated from the space-time foam should not be very simple. At last, the entropy outflow from an extra space to the main space should be supplied.

\subsection{Inverse landscape}

Landscape concept was introduced to describe the idea of the multiverse.  More definitely, starting from some primary Lagrangian with initial parameters, various universes with different properties are realized depending on initial conditions. Well known theory containing such possibility is the string theory. As was shown above, other ways also lead to the multiverse picture.

The concept of the Inverse Landscape is tightly connected with the Landscape ideology. It describes a set of primary theories that could lead to the observable low-energy physics.

Detailed study was pervormed in the paper \cite{2012GReGr..44.2283R} where it was shown that specific values of initial parameters are not very important provided that the multidimensional Lagrangian contains higher derivatives .
The essence of the idea is as follows. Consider a space of multiple dimensions. Because of quantum fluctuations in some of its regions, a geometry of
the direct product of two subspaces may arise. Suppose the curvature of one of the subspaces significantly exceeds that of the other; let us refer to the first
subspace as the extra one and to the latter as the main subspace. Quantum fluctuations in some region of a newly formed main subspace similarly divide it into direct product of two subspaces. Considering further divisions we arrive at a chain of partitions of the space. Multiple consecutive steps reduce effective dimensionality of the space. Every step of a cascade changes parameters of the Lagrangian.

Numerical simulations have indicated that there are numerous values of initial parameters of the theory that could be "connected by a cascade"\, with observed fundamental constants. Particular numerical values of initial parameters are therefore not as important as it was previously thought. The inverse landscape is the concept when the observed physics is derived from numerous initial theories. This inverted landscape model brings the significance of a search for the unique primary Lagrangian into question.

\section{Limits to a size of compact extra space}

In the framework of multidimensional gravity the low energy physical parameters strongly depend on extra space properties and on its size in particular. If the latter is small enough quantum fluctuations of its metric invalidate classical description. So we need some criterion of a "classicality"\, of compact space. This problem is discussed below.

Let we have action \eqref{act1} acting in a $(D-4)$-dim compact extra space. Its size must be greater than $1/m_D$ to avoid quantum fluctuations of a metric. This statement represents common point of view based on an analogy with the Einstein gravity.
In this Section it will be shown that the situation is more complicated and promising, see also \cite{2004GReGr..36..451W}.

Consider a  $D$-dim space $M_D = M_{D-4}\times M_4$ with a metric $G_{AB}$ and action
\begin{equation}
 S= \frac{m_D ^{D-2}}{2}\int d^D y \sqrt{G}R. 
\end{equation}
Let some quantum fluctuation of the metric takes place in the region $U$ with a characteristic size $l$ and a volume $l^D$. It leads to a fluctuation of the Ricci scalar $\delta R$ in a volume $\delta V_{D-4} \sim l^{D-4}$ of the extra space $M_{D-4}$ and in a volume $\delta V_4 \sim l^4$ of the main space $M_4$. A deviation of the action from its classical value may be estimated as
\begin{equation}\label{dS}
\delta S \sim \frac{m_D ^{D-2}}{2}\delta V_{D-4} \delta  V_4  \delta R .
\end{equation}
The probability of the metric fluctuation due to quantum effects is of order of unity if
\begin{equation}\label{dSless}
\delta S < 1.
\end{equation}

If a fluctuation of the Ricci scalar $\delta R$ in a region $U$ is comparable to the average value $\bar{R}$ of the Ricci scalar,
\begin{equation}
\delta R \sim \bar{R},
\end{equation}
then the classical description is not allowable. Using expressions \eqref{dS}, \eqref{dSless} one obtains a lower limit $l_{min}$ to a size of the region $U$ to which a classical description is applicable
\begin{equation}\label{QQU}
l>l_{min} \equiv m_D^{\frac{2-D}{D}}\bar{R}^{-1/D}.
\end{equation}

If a size $L$ of the $D-4$-dim extra space satisfies the condition
\begin{equation}\label{QQALL}
L<l_{min},
\end{equation}
then this extra space lies in the quantum region and hence belongs to space-time foam.
Let us apply expression \eqref{QQALL} to a compact $D-4$-dim extra space  of constant positive curvature, $\bar{R}=const >0$. In this case the connection between the Ricci scalar and the curvature radius
\begin{equation}\label{relati}
\bar{R}=n(n-1)/r_c^2
\end{equation}
is known. At the same time
the size of extra space is determined as
\begin{equation}
L=r_c
\end{equation}
and expressions \eqref{QQU},  \eqref{QQALL} give the lower limit to classical description in the form
 \begin{equation}\label{LmD}
L > m_{D} ^{-1}\left[n(n-1)\right]^{\frac{2}{D(2-D)}}\sim m_D ^{-1}.
\end{equation}
in full agreement with our expectations. An extra space size must be much larger than the inverse $D$-dim Planck mass for its classical description.

Geometry of compact hyperbolic spaces is more complicated. Instead of relation \eqref{relati} we have, see\cite{2002PhRvD..66d5029N}
\begin{equation}
\bar{R} = n(n-1)/r_c ^2, \quad  V_{D-4}\simeq r_c ^{n} e^{\alpha};\quad \delta V_4 \sim L^4
\end{equation}
where $r_c$ is the curvature radius of the compact  hyperbolic  space and
\begin{equation}
\alpha = (n_{eff}-1)L/r_c , \quad L\gg r_c .
\end{equation}
There is no strict connection between a curvature radius of extra space and its size.
Together with expressions  \eqref{dS}, \eqref{dSless} these estimates give
\begin{equation}\label{neq1}
(m_D r_c )^{D-2}>\frac{(n_{eff}-1)^4    }{n(n-1)}\alpha^{-4} e^{-\alpha}.
\end{equation}

Finally the condition for classical description of a compact hyperbolic space may be written in the following form
\begin{equation}\label{rmDhyp}
r_c >r_{min} m_D ^{-1}, \quad r_{min} =\left[\frac{(n_{eff}-1)^4    }{n(n-1)} \alpha^{-4} e^{-\alpha}\right]^{1/(D-2)}.
\end{equation}
This condition is less restricted than that in \eqref{LmD}. As an illustration, consider $2$-dimensional extra space with the size $L =10^{-2} m^{-1}_D$. The case of negative curvature is represented in Fig. 1 where the classical condition \eqref{rmDhyp} is fulfilled for small curvature radius even if $L$ much smaller than $1/m_D$.
The unexpected feature is that the smaller the curvature radius is, the better condition \eqref{rmDhyp} is satisfied and a hyperbolic compact space with the same size $L$ becomes "more classical".

The discussion above indicates that a production of compact spaces with various properties is an important ingredient of future theory. The last Section devoted to direct observational effect of such production.

\begin{figure}
\centering
\includegraphics[width=0.7\linewidth]{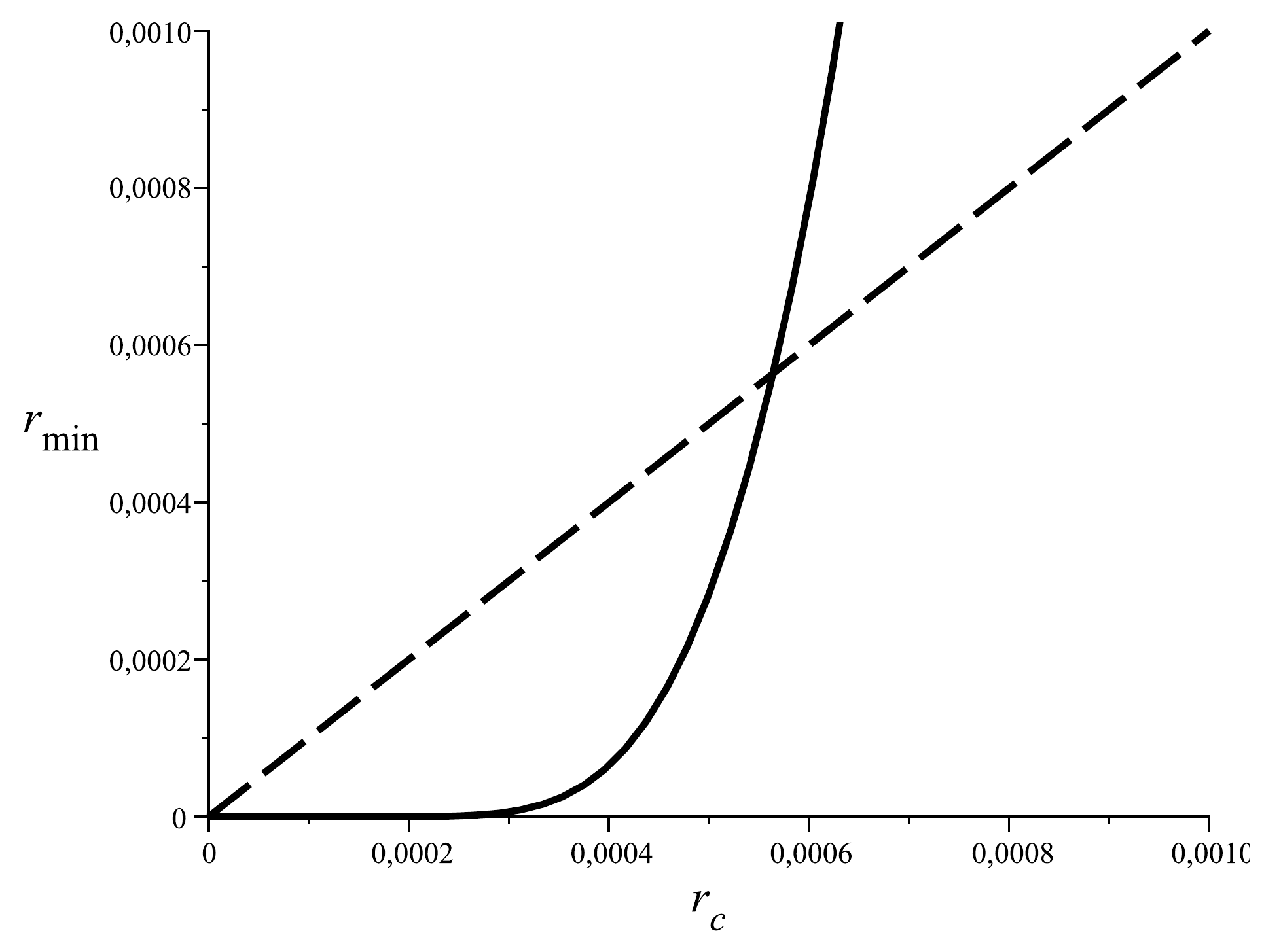}
\caption{The curvature radius $r_c$ (dashed line) must be larger than minimal curvature $r_{min}$ (solid line) to suppress quantum effects. Numerical parameter values - $n=n_{eff}=2$, the size of extra space $L=0.01$ in $m_D$ units.}
\label{fig:plot}
\end{figure}

\section{Inhomogeneous states}

It is known that an extra space geometry can be responsible for the dark matter  \cite{2008PhRvD..77k5012P,2009MPLA...24..667K}. In this Section we shortly discuss the subject.
As was shown above properties of extra space depend on initial conditions. Consider some region $U_+ \subset $ $M_4$ with a size $l_+$. Let each point of this region be equipped with $2$-dim extra space $M_+$ with the Ricci scalar $R=R_+ =const$.  The rest of $4$-dim space $M_4\setminus U_+$ is equipped with $2$-dim extra space $M_{torus}$ with the metric \eqref{torus1}. This configuration is stable due to difference in the topologies of extra spaces $U_+ $ and   $M_4\setminus U_+$.

The energy density of the main space $M_4\setminus U_+ $ may be expressed in the following way
\begin{equation}
\rho_{torus}=-\frac{m_D ^{(D-2)}}{2}\int d^2 y \sqrt{G(y)} f(R_{torus})
\end{equation}
where formulas \eqref{density}, \eqref{Rtor} have been used.
In a similar manner the energy density of the region $U_{+}$ has the form
\begin{equation}
\rho_{+} =-\frac{m_D ^{(D-2)}}{2}\int d^2 y \sqrt{G(y)} f(R_+ ) \simeq -m_D ^{(D-2)} V_2 f(R_+ )
\end{equation}
and the mass of this region may be estimated as
\begin{equation}
M\simeq l_+ ^3 (\rho_{torus}-\rho_{+})
\end{equation}
for a distant observer.

Physical laws within such areas differ from ordinary ones. Could they be found somehow, except due to gravitational effects? It would be interesting to investigate what happens to an object, such as a star flying into the region $U_+$.

It is natural to assume that the minimum size of such a region is of the order of the extra space size, i.e. smaller than $10^{-18}$cm. Such particle-like objects (WIMPs) interacts only gravitationally with ordinary particles and can constitute the dark matter.

Remarkable feature of such particles is that any two of them can have different masses. Indeed, the first period of the Universe formation - inflation - is characterized by quantum fluctuations. They could have produced different initial conditions in causally disconnected space regions. It means that the extra space metric and hence masses of the particle-like objects  in such regions also differ from each other.

\section{Conclusion}

It was shown in this paper that the pure multidimensional gravity contains reach set of mechanisms describing low energy physics.
General picture is as follows. Firstly, some Riemann manifold is formed from the space-time foam. Secondly, a stage of classical evolution takes place and its result depends on both initial metric of the manifold and a proposed Lagrangian. The third epoch consists of the extra metric stabilization due to the entropy outflow into the main space. Classical size of a compact extra space depends on a form of function $f(R)$ in \eqref{act1} and initial conditions.

The separate problem is a minimal size of classical extra space. Common point of view is that its curvature radius must be greater than the D-dimensional Planck mass. As was shown above this restriction is much weaker in case of the compact hyperbolic extra spaces.

Initial metric created from quantum space-time foam influence not only the low energy parameters but is also responsible for a group structure of gauge fields. Two ways of the simplest $U(1)$ symmetry formation are considered. The first one is torus topology of extra space which arises in case of the specific form of Lagrangian.
The second possibility takes place due to point-like defects of metrics. Continuous set of such compact 2-dimensional metrics has similar behavior near $\theta = 0$. This opens a door for the concept of continuum landscape.

As was shortly discussed in last section such topological and/or point-like defects could be observed as dark matter particles (WIMPS). This subject is worth studying separately.

\textbf{Acknowledgment}

The author is grateful to K. Belotsky, A. Bolozdynya, R. Konoplich, M. Skorokhvatov and I. Svadkovsky for their interest and fruitful discussions.
This work was supported by grant rscf 14-42-00065.


\end{document}